# Visualization and Characterization of Agricultural Sprays Using Machine Learning based Digital Inline Holography

*Shyam Kumar M [1], Christopher J. Hogan [1], Steven A. Fredericks [3], Jiarong Hong [1,2]*

[1]Department of Mechanical Engineering, University of Minnesota, Minneapolis, MN, USA

[2]Saint Anthony Falls Laboratory, University of Minnesota, Minneapolis, Minnesota, USA

[3] Winfield United, River Falls, Wisconsin, USA

Corresponding author: Jiarong Hong, PhD
　　　　　　　　　　　Professor
　　　　　　　　　　　Department of Mechanical Engineering
　　　　　　　　　　　University of Minnesota
　　　　　　　　　　　Phone: 612-626-4562
　　　　　　　　　　　Email: *jhong@umn.edu*

**Abstract**

Accurate characterization of agricultural sprays is crucial to predict in field performance of liquid applied crop protection products. In this paper, we introduce a robust and efficient machine learning (ML) based Digital In-line Holography (DIH) algorithm to accurately characterize the droplet field for a wide range of commonly used agricultural spray nozzles. Compared to non-ML based DIH processing, the ML-based algorithm enhances accuracy, generalizability, and processing speed. The ML-based approach employs two neural networks: a modified U-Net to obtain the 3D droplet field from the numerically reconstructed optical field, followed by a VGG16 classifier to reduce false positives from the U-Net prediction. The modified U-Net is trained using holograms generated using a single spray nozzle (XR11003) at three different spray locations; center, half-span, and the spray edge to create training data with different number densities and droplet size ranges. VGG16 is trained using the minimum intensity projection of the droplet 3D point spread function (PSF). Data augmentation is used to increase the efficiency of classification and make the algorithm generalizable for different measurement settings. The model is validated using National Institute of Standards and Technology (NIST) traceable glass beads and six different agricultural spray nozzles representing various spray characteristics. The principal results demonstrate a high accuracy rate, with over 90% droplet extraction and less than 5% false positives, regardless of droplet number density and size. Compared to traditional spray measurement techniques, the new ML-based DIH methodology offers a significant leap forward in spatial resolution and generalizability. We show that the proposed ML-based algorithm can extract the real cumulative volume distribution of the NIST beads, where the LD system measurements are biased towards droplets moving at slower speeds. Additionally, the ML-based DIH approach enables the estimation of mass and momentum flux at different locations and the calculation of relative velocities of droplet pairs, which are difficult to obtain using conventional spray characterization techniques.

# 1. Introduction

Sprays are widely used in different industrial systems, including in internal combustion engines, spray coating, fire suppression, and spray towers, alongside agriculture applications. Typically, the purpose of spray implementation is to increase the interfacial area and the coverage or exposure of the applied liquid to the system of interest. In particular, crop protection products are applied via sprays because sprays enable cost-effective application at agronomically relevant timings with uniform coverage throughout the field. Although dispersing tank mixtures into droplets leads to improved coverage, droplets can be entrained by prevailing atmospheric conditions, and transported off-target, leading to spray-drift and a reduction in real application rate (Holterman et al. 1997). Raising application rates to counteract spray drift not only inflates the associated costs but can also precipitate harmful effects on neighboring crops, surface water, ecosystems, and human health, as noted by Baetens et al. (2007). Given these implications, the control of spray droplet size distributions emerges as an essential measure for maximizing grower outcomes and safeguarding both our agricultural practices and the surrounding environment. The drift control measure is primarily intended to shift the droplet size distribution of the spray to larger diameters, by which the production of drift-prone smaller droplets can be reduced (Nguyen et al. 2023).

The size distribution of sprays is influenced by the physical properties of the tank mix (Makhnenko et al. 2021, Nguyen et al. 2023), the nozzle (Guler et al. 2007) and the nozzle operating conditions (Kooij et al. 2018). Therefore, when measuring the drift potential of nozzle-tank mix combinations, consistent methods of quantifying spray droplet size distributions, i.e., the relative volumetric concentration of droplets as a function of size, are frequently used. Temporal, one-dimensional measurement techniques, such as phase Doppler interferometry (PDI) (Kumar et al. 2019a) and laser diffraction (LD) (Sijs et al. 2021), are widely utilized for droplet size distributions (DSD) measurements. These techniques offer high temporal resolution and relatively fast measurement capabilities, making them well-suited for capturing dynamic changes over time. They are applicable to both sparsely concentrated and dense regions within sprays. However, they often have lower spatial resolution, and the data inversion techniques employed with them may lead to difficulties in measuring non-spherical droplets and in instances where multiple droplets are detected simultaneously (Privitera et al.2023). Conversely, spatial measurement techniques such as shadowgraphy, particle image velocimetry (PIV), and laser-induced fluorescence (LIF) (Clemens 2002) offer high spatial resolution and direct measurements, reducing the need for extensive sampling. However, spray behavior is three-dimensional, and many spatial measurement techniques have a limited depth of focus. Detecting very fine droplets poses a challenge as well, and in dense spray regimes, multiple light scattering can lead to misleading results (Mishra et al. 2017). Finally, while tomographic imaging and 1p-2p (Mishra et al. 2017) methods show promise in addressing some of these limitations, they remain under development towards commercial (i.e., non-research) applications. Overall, there is presently a need for improved, cost-effective, accurate, and spatially sensitive DSD measurements in agricultural sprays.

High-resolution digital inline holography (DIH) is a direct imaging approach of rapidly growing interest in analyzing droplet and particle dynamics in flows (Katz and Sheng 2010). DIH has several advantages over aforementioned techniques, such as high temporal and spatial resolution, low-cost, highly compact setup, and large depth of focus. DIH has been extensively utilized in various spray settings to examine the behavior and characteristics of droplets (Gao et al. 2013, Kumar et al. 2019b, and Wang et al. 2022). Our research team has recently highlighted the utility of DIH in the field of agricultural spray DSD analysis, as documented in our studies from Kumar et al. (2019b) and (2020). Furthermore, we have illustrated the ability of DIH to extract multidimensional data, specifically, bivariate size-velocity distributions for droplets (Li et al. 2021). These investigations, however, were constrained to a singular spray type and a limited droplet size range. The hologram processing methods employed in these studies encompassed several stages, including reconstruction, focus detection, and segmentation. While these traditional methods have proven effective, they inherently limit robustness, generalizability, and processing speed. Additionally, in order for the conventional hologram processing method to accurately segment a hologram, it necessitates meticulous fine-tuning of parameters to achieve precise results. This involves adjusting the threshold value for segmentation and selecting appropriate focus metrics. The complexity of fine-tuning increases when dealing with a large dynamic range of particle sizes. These limitations present substantial hurdles for the broader implementation of these techniques in commercial and industrial environments. In our current work, we strive to overcome these challenges and build upon our previous research by incorporating machine learning (ML) based post-processing techniques into DIH. This progressive approach amplifies the applicability and efficiency of DIH, rendering it a more feasible choice for commercial and industrial use. The use of ML-based DIH can significantly contribute to the potential for in-depth spray characterization, offering enhanced product development for agricultural sprays. This approach can capture spatial and temporal variations in DSD, potentially address unsteadiness in sprays, and map the distribution of different droplet classes across the spray span with high accuracy and reduced human effort. These insights could play a crucial role in establishing new benchmarks for reduced drift spray nozzles, promoting precision in application and enhanced environmental safety.

Our ultimate objective is to devise a robust spray analysis methodology and demonstrate its precision and generalizability in characterizing spray data pertinent to agricultural sprays, utilizing standard reference nozzles (ISO 2018, ASABE 2020). In the sections that follow, we first provide an overview of DIH and a brief description of prior-post processing efforts to justify the need and utility of ML based approaches. We then describe the specific experiments performed to develop, validate, and ultimately apply DIH with ML. Finally, we confirm the accuracy of this approach with National Institute of Standards and Technology (NIST) traceable silica beads and determine DSDs for the six ASABE boundary reference nozzles ranging from very fine/fine to extra-coarse/ultra-coarse.

## 2. Materials and Methods
### 2.1 Digital Inline Holography Overview

The standard optical arrangement of a DIH system (Fig. 1a) consists of a laser source, collimating optics, and a camera with an imaging lens (Toloui and Hong 2015, Toloui et al. 2017). The coherent light passes through a spatial filter to increase spatial coherence, followed by a collimator lens. The resulting expanded, highly collimated beam, known as the reference light, is used to illuminate the particle field. Each particle (i.e., droplet, referred to as a particle here to be consistent with common terminology in applying DIH) in the field scatters the laser light, creating object light. The interference between the reference and object light produces hologram patterns as a 2D aperture on the camera sensor. Fig. 1(b) displays a sample hologram recorded for a particle field.

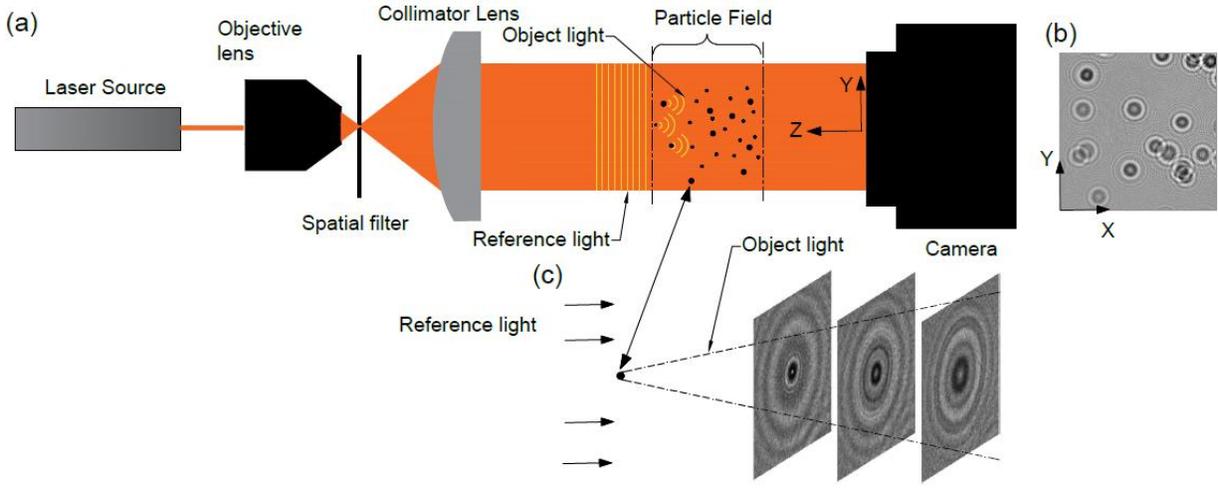

*Fig. 1*: *(a) Schematic of optical setup of DIH (b) Representative hologram (c) Representative schematic of interference pattern of a particle formed at different Z locations.*

It is important to note that the interference patterns overlap when particles are closer in the lateral XY plane. Unlike other flow visualization techniques, DIH records both the phase and intensity information of each particle. Fig. 1(c) illustrates the interference patterns of a particle at different longitudinal locations. The patterns are distinct for different Z-locations. This phase information can be digitally reconstructed from the recorded hologram using numerical algorithms such as Fresnel transforms. In DIH, the complex 3D optical field is can be reconstructed as (Berg 2022):

$$u_p(x, y, z) = u(x', y') \otimes h(x', y', z) \quad (1)$$

where $u_p$ is the reconstructed 3D optical field intensity, $x$, $y$ and $z$ are lateral and longitudinal locations respectively, $x', y'$ represents the lateral locations in the imaging plane (camera sensor), $u$ is the recorded hologram, $\otimes$ represents the convolution operator and $h$ is the point-spread function (PSF), which describes how a point source is spread or blurred in the recorded holograms due to the properties of the optical setup used. The corresponding Fresnel transform can be written as:

$$u_p(x, y; z) = \frac{exp(i2\pi z/\lambda)}{i\lambda z} \iint_{-\infty}^{\infty} u(x', y') exp\left\{\frac{i\pi}{\lambda z}[(x - x')^2 + (y - y')^2]\right\} dx' dy' \quad (2)$$

where $u_p$ is the reconstructed 3D optical field and $\lambda$ is the wavelength of the laser beam. The resulting reconstruction field contains the 3D intensity field of all the particles. Different algorithms can be used to identify each particle's 3D position and other characteristics (Kumar et al. 2019b, Berg 2022).

The post-processing of holograms generally involves three main steps. First, pixels with a high probability of containing droplets are identified. Subsequently, a focusing technique is employed to precisely determine the focal plane within the reconstructed 3D optical field. Finally, a segmentation approach is utilized to detect and analyze the sizes of droplets in the hologram by dividing it into distinct regions as droplets and the rest of the optical field. Non-ML based DIH methods often employ thresholding to segment images into particles. However, the threshold values can vary, depending on the different conditions under which holograms are acquired, and are often selected manually. Gao et al. (2013) introduced a hybrid algorithm that utilizes the minimum intensity projection for droplet identification. It then employs a pixel intensity gradient (sharpness) based focus metric to ascertain the depth of each particle. Each in-focus grayscale image is subjected to an iterative thresholding method for segmentation. Kumar et al. (2019b, 2020), Li et al. (2021), and Wu et al. (2021) have adopted similar approaches. The use of a wavelet filter on the reconstruction field to obtain the focus metric is also widely used (Wang et al. 2022). However, while using minimum intensity projection and wavelet filters are most effective for larger particles, the use of pixel gradient can compromise the sharpness measurement of smaller particles. To mitigate noise interference, particles with a pixel number exceeding a certain threshold value are typically considered as particles (Wang et al. 2022). In summary, the segmentation methods used in non-ML DIH often require human intervention for fine-tuning parameters to achieve optimal performance, particularly when dealing with holograms captured under varying conditions. This necessity for manual adjustment limits the generalizability of the method used.

The incorporation of machine learning (ML) into hologram processing presents a compelling solution to transcend the constraints associated with thresholding methods. This integration not only enhances the generalizability and precision of droplet analysis across diverse spray settings, but also significantly boosts the processing speed, thereby streamlining the overall analysis process. Shao et al. (2020) has demonstrated the use of a modified U-Net architecture on spray hologram segmentation to detect spray droplet positions and sizes. ML based spray histogram segmentation is found to be more effective than the non-ML based conventional method (Shao et al. 2020). However, the droplet extraction rates are diminished for higher number density polydisperse droplet-fields (false negatives). Conversely, for low-number density sprays, the false positive rate is found to be higher because of the class-imbalance problem. Chen et al. (2021), has considered holograms as a compressive sensing problem and used iterative shrinkage thresholding where the parameters are automatically tuned using an unrolled neural network. However, this method needs 3D training labels, which increases the GPU memory usage and forces use of a small region of interest. Further, extensive studies that show the generalization of the ML based spray hologram segmentation is limited.

*2.2 Spray Measurements*

The experiments for the current study were conducted in a closed recirculating wind tunnel described in detail previously (Kumar et al. 2019b, Nguyen et al. 2023). A top-down view schematic diagram of the test section of width 0.91 m, height 1.83 m and length 3.20 m is shown in Fig. 2 (a). The spray nozzle is mounted on a stepper-motor-controlled one dimensional (*y*-axis) traverse. The wind tunnel test section side walls are made of glass to provide optical access for the DIH system. A 50 mm diameter collimated laser beam, represented using red in Fig. 2a, is produced using a 12 mW He-Ne 632 nm Laser source (REO Inc.), a neutral density filter, objective lens, a 10 µm spatial filter (Newport Inc.) and a 75 mm focal length collimation lens (Thorlabs Inc.). The imaging system consist of a high-speed camera (Phantom v710) of spatial resolution of 18.2 µm/pixel and a 105 mm imaging lens (Nikon 105 mm f/2.8).

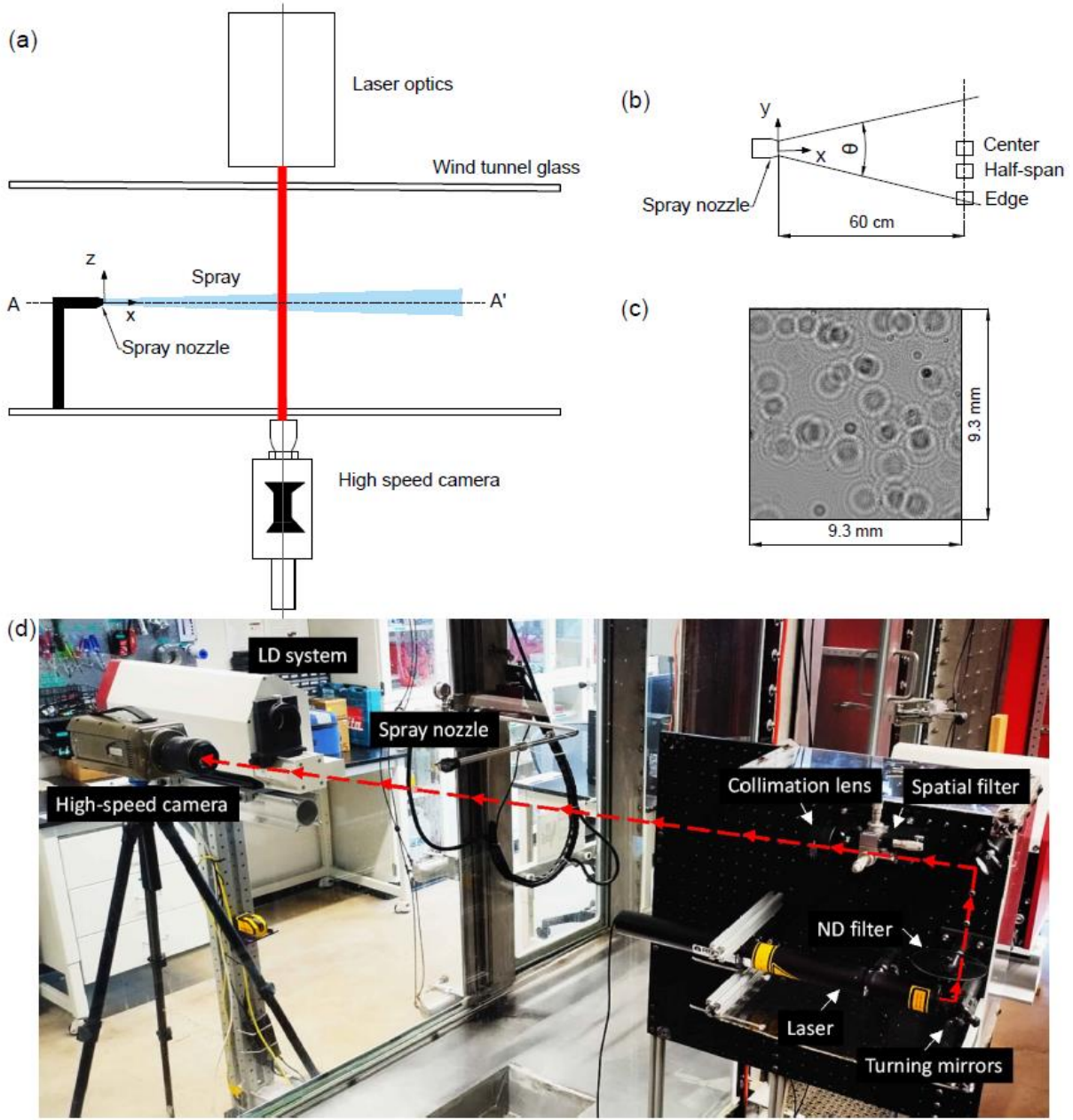

*Fig. 2. (a) Schematic diagram of the experimental setup employed. The axial and longitudinal directions are specified by the x- and z-axes, respectively, with the origin placed at the spray nozzle exit. (b) Front view of the AA' plane indicated in (a), along with the holography measurement locations. (c) A representative hologram obtained using DIH over a 9.3 mm × 9.3 mm window. (d) A photograph of the DIH imaging setup. The Laser direction is shown in red dashed lines.*

Flat fan nozzles producing polydisperse droplets were used for measurements. To define the position of the measurements within these sprays, the origin is defined at the nozzle exit with $x$ and $z$ axis along the longitudinal and lateral directions, respectively. The front view of section AA' is shown in Fig. 2(b). DIH measurements were performed at three different y locations, i.e., center, half-span, and edge, at $x = 60$ cm downstream. To make sure that the DIH measurement regions are physically similar across different nozzles, center, half-span and edge are defined as $y = 0\ cm$, $-60tan(\theta/4)\ cm,\ -60tan(\theta/2)\ cm$ , respectively, where $\theta$ is the nozzle spray angle (see Fig. 2b). For each measurement location, images were acquired at 25000 frames per second, with 512 × 512 pixel$^2$ on the camera representing a field of view of 9.3 × 9.3 mm$^2$. A representative hologram recorded using the current optical setup is shown in Fig. 2(c). These holograms are further postprocessed for estimating droplet 3D locations and diameters.

The details of the spray nozzles used to acquire the holograms are noted in Table 1. Seven TeeJet nozzles (TeeJet Technologies, Glendale Heights, IL) were used in the study: one XR11003 (used in ML training) and six ASABE reference nozzles: TP11001-SS, TP11003-SS, TP11006-SS, TP8008-SS, TP6510-SS, and TP6515-SS corresponding to the spray classification boundaries: Very Fine/ Fine (VF/F), Fine/Medium (F/M), Medium/Coarse (M/C), Coarse/Very Coarse (C/VC), Very Coarse/Extremely Coarse (VC/XC) and Extremely Coarse/Ultra Coarse (XC/ UC), respectively (ISO 2018, ASABE 2020). The nozzles and pressure combination define the boundaries between size classifications and are used herein as a proxy for generating all relevant droplet size distributions for the majority of agricultural applications. The XR11003 body is made of acetal polymer, and the orifice is made of stainless steel. The reference nozzles, however, are entirely made of stainless steel. In all cases the spray was water, operated at the nozzle pressure noted in Table 1. When measuring these sprays, the wind tunnel was operated at a speed of 2.68 m/s.

*Table 1. Details of the spray nozzles used in the current study. The nozzle ID of the six reference nozzles, their operating pressures and spray angles are based on spray category thresholds, as defined by American Society of Agricultural and Biological Engineers (ASABE 2020). The corresponding flow rate and the droplet sizes (D0.1, D0.5, D0.9) are measured using the LD system in the wind tunnel as shown in Fig. 2c. D0.1, D0.5 and D0.9 are the average droplet diameter along the entire span of the spray at 10%, 50% and 90% of the cumulative distribution.*

| Nozzle types | Nozzle ID | Nozzle pressure (bar) | Spray angle (deg) | Flow rate (lpm) | D0.1 (μm) | D0.5 (μm) | D0.9 (μm) |
|---|---|---|---|---|---|---|---|
| XR11003 | XR11003 | 2.06 | 110 | 0.94 | 133 | 205 | 356 |
| Reference nozzles | TP11001-SS | VF/F | 4.50 | 110 | 0.41 | 65 | 139 | 273 |
| Reference nozzles | TP11003-SS | F/M | 3.00 | 110 | 1.16 | 101 | 227 | 390 |
| Reference nozzles | TP11006-SS | M/C | 2.00 | 110 | 1.93 | 135 | 320 | 498 |
| Reference nozzles | TP8008-SS | C/VC | 2.20 | 80 | 2.74 | 150 | 357 | 547 |
| Reference nozzles | TP6510-SS | VC/XC | 1.20 | 65 | 2.52 | 194 | 436 | 649 |
| Reference nozzles | TP6515-SS | XC/UC | 1.00 | 65 | 3.46 | 274 | 645 | 1024 |

*2.3 Machine Learning Approach*

For hologram segmentation, Shao et al. (2020) used a U-Net based architecture with a three-channel input and a two-channel output. The three-channel input includes the enhanced hologram, reconstructed 3D optical field, and minimum intensity projection. The enhanced hologram is determined by subtracting the background image (estimated using the whole set of acquired holograms) from the raw hologram. The reconstructed 3D optical field is estimated using equation (2) and the minimum value of the reconstructed 3D optical field along the z axis at each (*x, y*) location is represented as the minimum intensity projection (see Fig. 3). The output channels are binary stacked images of particle size and centroid for different reconstruction planes. Shao et al.

(2020) demonstrated that this proposed ML based segmentation is efficient in particle detection towards flow visualization. However, there are some limitations to this ML model in spray analysis. First, the particle detection rate reduces with increasing the droplet number density. Second, after segmentation, a manual thresholding is required to reduce the false positive rate, which is reasonable for particle fields containing larger particles. Conversely, for spray fields with high number density and small mean diameter this thresholding can filter out weak signals from small droplets. To avoid these limitations, a modification on the U-Net method by Shao et .al (2020) is implemented here.

We used two central neural networks (CNNs), one for segmentation and the second for classification. A detailed illustration of the architecture used is shown in Fig. 3. For hologram segmentation, we again employ the U-Net architecture proposed by Shao et al. (2020), but with a modification to accommodate the detection of droplets irrespective of size and number density. The schematic illustration of the Modified U-Net used for the current study is shown in Fig. 4.

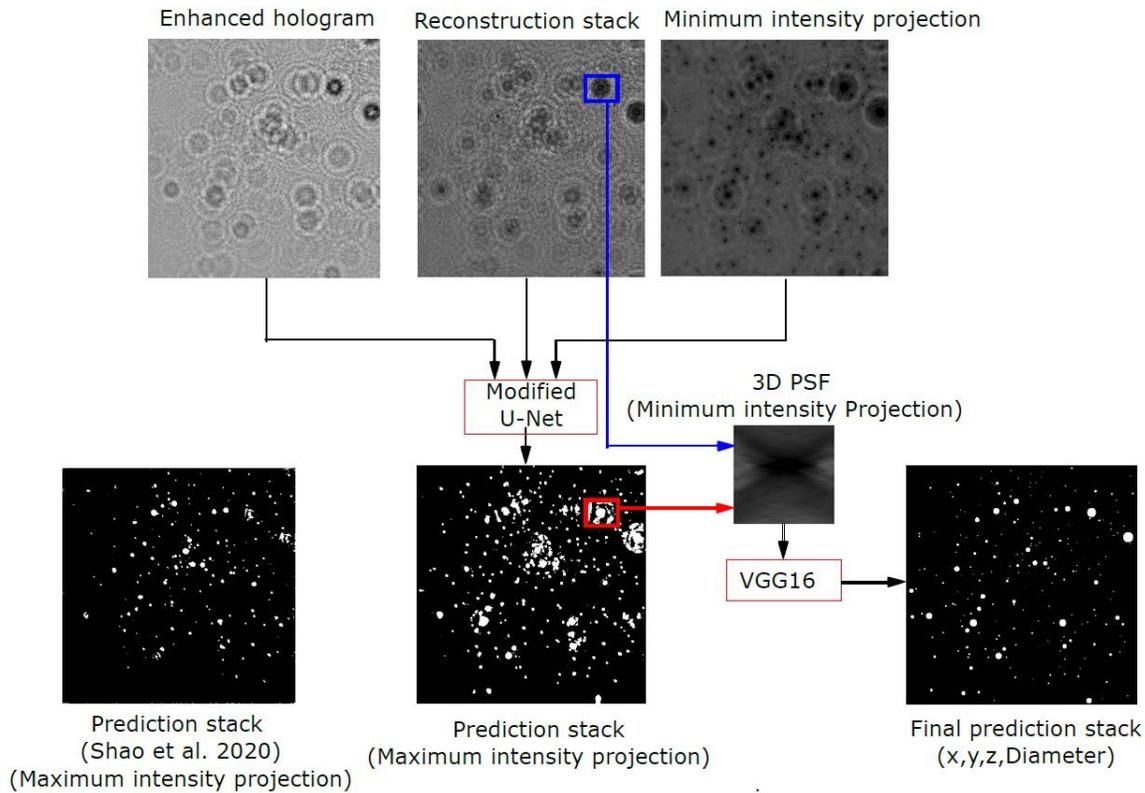

*Fig. 3*: *Detailed flowchart illustrating the model employed for droplet detection and segmentation. The prediction stack obtained using Shao et al. (2020) is also shown for comparison.*

The modified U-Net consists of an encoder, densely connected convolutions, and a decoder. The encoder comprises three encoder blocks, each consisting of two 3x3 convolutional filters with ReLU activation, followed by 2x2 max pooling (Fig. 4). The number of feature maps is doubled in each block, increasing the number of layers at each step. Following the encoder path, densely connected convolutions are employed. In densely connected convolutions, feature maps learned from all previous layers are concatenated with the feature map learned from the current layer (Azad et al. 2019). The decoding path incorporates three decode blocks, each consisting of up-sampling followed by batch normalization and ReLU activation. The output of the batch normalization is usually concatenated with the corresponding features from the encoder path using skip connections. Inspired by bi-directional convolutional long short-term memory (BConvLSTM) (Azad et al. 2019), the U-Net was further modified by using BConvLSTM in the skip connection rather than simple concatenation, to yield feature maps that are rich in local and sematic information. By the end of the decoding path the size of the feature maps reaches the original size of the input image (see Fig. 4).

Two separate output layers are defined, each for the size and centroid location, which uses a convolution with a sigmoid activation function. For this modified U-Net, to improve generalizability a 50% neuron dropout along with modified skip connection was applied to specific convolution layers in the encoder part of the U-Net (Azad et al. 2019). In the context of neural networks, a 50% neuron dropout refers to randomly deactivating (setting to zero) 50% of the neurons in a layer during the training phase. By randomly dropping out neurons, the network is driven to learn more robust and diverse representations. Additionally, dropout helps regularize the feature maps at different stages of U-Net, reducing the risk of overfitting. In total approximately 94 million neurons are used in this model.

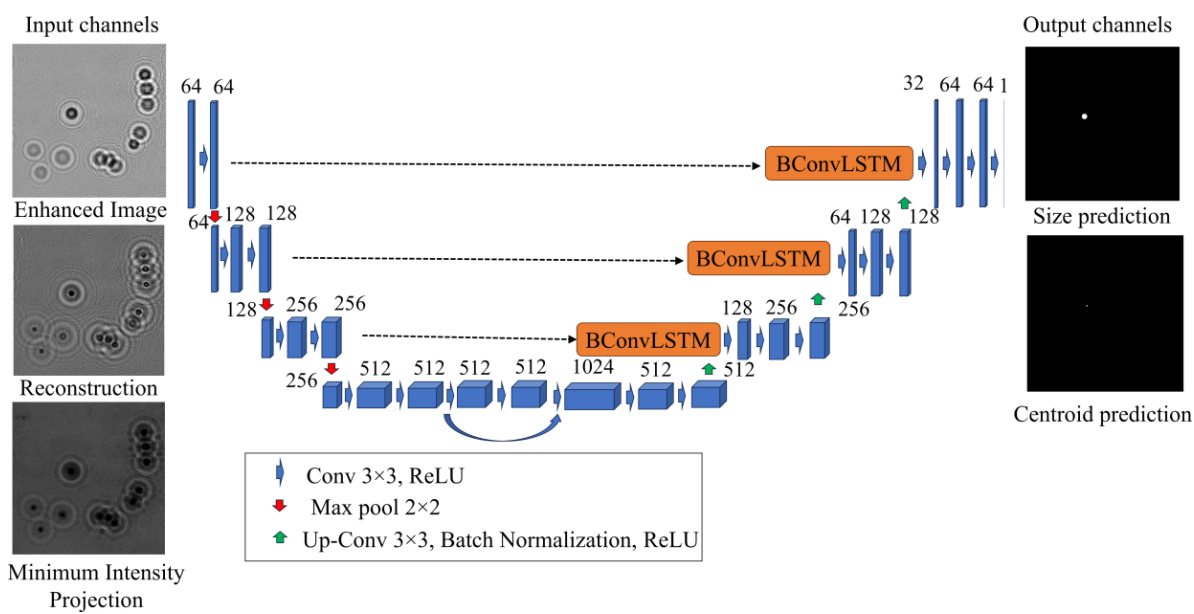

***Fig. 4***: *Schematic of the architecture of the modified U-Net. The number of filters/feature map corresponding to each step is also shown.*

To train the U-Net, holograms were acquired using XR11003 nozzles operating at 2.06 bar. To improve the generality of the training set holograms were included from multiple measurement locations, with a total of 150 spray holograms from the center, half-span, and edge locations, randomly split into training and validation set with an 80:20 ratio. Each hologram was numerically reconstructed into 251 planes for a total of 37650 distinct images used in training and validation. It is worth noting that all the images (512 x 512 pixel$^2$) were reshaped to 256×256 pixel$^2$ images for the training process to improve memory usage. The training targets consist of two channels, the particle size and centroid. If a reconstruction plane contains any in focus droplets, the corresponding training target contains the binary image of the droplet, with the droplet in white and background in black, and the corresponding centroid location. The details of the training labels are elaborated in the supplementary materials. The predictions for the training images were generated using the U-Net proposed by Shao et al. (2020). To refine the labels, an in-house built graphical user interface (GUI) is utilized for manual updates, including removing false detections and adding undetected droplets. These updated labels then serve as the training targets for the modified U-Net. Binary cross entropy is used as the loss function for both channels. The modified U-Net architecture was implemented using Keras with the Tensorflow backends with the training conducted on a Nvidia RTX 2080Ti GPU. The Adam optimizer is used with a learning rate of 0.0001. The model was trained for 30 epochs. The holograms were shuffled randomly during the training process, however, during prediction consecutive reconstruction planes were used to produce the predicted 3D droplet field that contains the size and centroid map. Fig. 3 shows the maximum intensity projection of the predicted 3D droplet field generated using the modified U-Net for a single hologram. While the real predicted 3D droplet field is in gray scale, to demonstrate effective detection, the gray scale is converted to binary image in Fig. 3. In Fig. 3, the prediction generated using U-Net proposed in Shao et.al (2020) is shown for comparison. By manual inspection, smaller droplets that were difficult to detect using the previous U-Net are found to be detected using this modified U-Net.

Alongside improved detection, a significant increase in the number of false positives was also observed. The false positives include signals from the interference rings of larger droplets and signals resulting from the overlapping of interference rings of nearby droplets. As noted above, previous studies used thresholding or a minimum number of pixels to avoid these false positives (Gao et al. 2013, Wang et al. 2022). However, these approaches may filter out smaller droplets, hence we used a second CNN, the well-developed VGG16 binary classifier (Simonyan and Zisserman 2014), to differentiate between droplets and noise.

From our preliminary analysis, the extracted 3D Point-Spread Function (PSF) from the reconstructed 3D optical field corresponding to droplets is clearly distinguishable from noise that would produce a false positive. This feature of the 3D PSF was isolated by creating a bounding box around the projected maxima of the prediction (shown as a red box in Fig. 3) and reconstructing the 3D optical field along its *z* axis from the reconstruction stack (represented by the blue line in Fig. 3). This was then collapsed along the *y* direction to produce a 224 x 224 $\text{pixel}^2$ 2D prediction maxima which can be input in VGG16.

When training the VGG16 model, the minimum intensity of 3D PSFs of multiple droplets were generated from real holograms. Similar steps were used to estimate the 3D nature of the false positives as well. To increase the efficiency of classification and to make the algorithm generalizable for different measurement settings, data augmentation was used. These augmentations included intensity variation (light and dark backgrounds), Gaussian noise addition and orientation (rotation, flip and mirror) and were used to generate 70,000 training images of droplets and false positives. The VGG16 model was trained using the same system specifications that were used when training the U-Net. The Adam optimizer and a sparse categorial cross-entropy loss function were used and trained for 50 epochs.

Diameter estimation using the predicted 3D droplet field, which has some size disparities, can introduce significant errors in the flow visualization and droplet sizing. To detect the z-location of each detected droplet, Laplacian filter-based focus metrics were employed on the reconstructed 3D optical field. Additionally, a 1D Gaussian filter and a Sobel filter were utilized to extract the droplet shape at the focal plane. These processes resulted in the generation of the final predicted 3D droplet field (as shown in Fig. 3), consisting of binary images representing droplets at their respective focal planes (z-locations). From the binary output, the details of the individual droplets are extracted using a feature extraction module (Van der Walt et al. 2014), which facilitated the computation of attributed such as area, centroid, bounding box etc. Droplets are predicted as white labels on their respective planes against a black background. Because the droplets are distributed across various planes, each is distinctly identifiable, contrasting the overlapping that can occur with single image projections when the number of droplets is high. Details of each droplet (3D location, diameter) are extracted to a text file, which is used for further analysis.

## 3. Results
### *3.1 Model Validation: Bead Measurements*

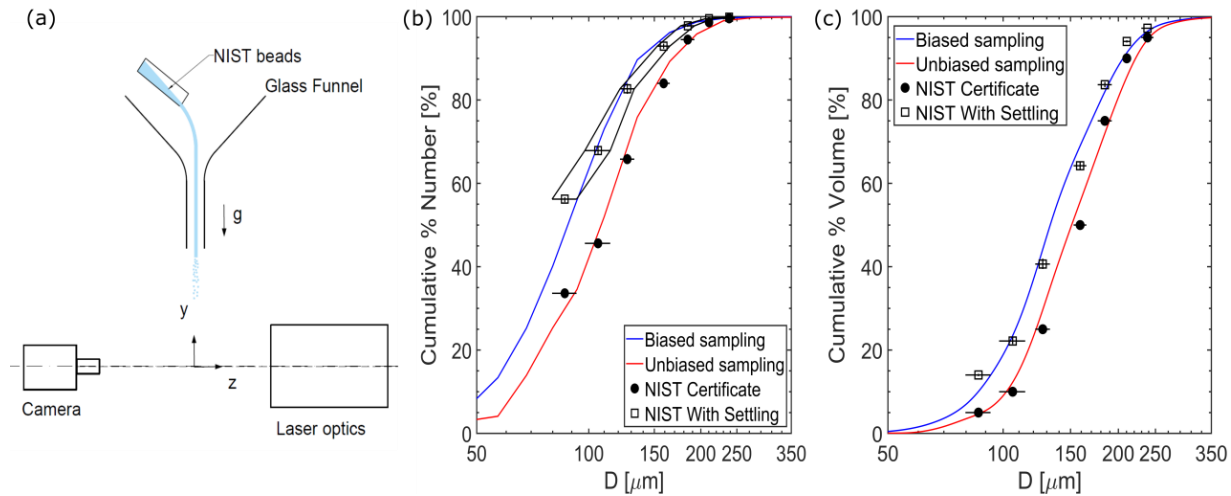

*Fig. 5. Validation of the model using NIST beads. (a) Schematic representation of the NIST beads-based experiment. Cumulative number distributions (b) and volume distributions (c) of NIST beads detected using the proposed DIH-based method for both biased and unbiased sampling. The NIST distribution provided, and modified for settling-based sampling biased are also plotted.*

Prior to presenting and analyzing results for spays, in model validation, we first present results for the measurement polydisperse glass microbeads with a NIST traceable size distribution in the 50-350 µm range (Whitehouse Scientific). Using the method described by Fredericks (2020) for laser diffraction calibration, a funnel is vertically aligned to the axis of the wind tunnel and the beads are poured from a height through the funnel as shown in Fig. 5(a). The recorded holograms as the beads gravitationally fall through the measurement volume are post-processed using the proposed model. The resulting cumulative number and volume distributions are presented in Fig. 5(b) and (c), respectively. These plots are generated for two sampling methods. When the holograms are consecutive in time or separated by only by a few time steps (less than 2 ms), there can be a bias toward smaller particles because of diameter resolved differential settling (biased sampling, the blue line) due to smaller particles have slower settling velocities than larger particles. On the other hand, when each hologram is separated by ~4 ms this bias can be avoided (unbiased sampling, the red line). As described in Fredericks (2020), trajectory calculations can be used to transform the cumulative size distribution provided in by the NIST traceable standard (labelled as "NIST certificate", the circle points), to settling biased distribution by weighting the distribution by the inverse of settling velocities (labelled as "NIST with settling", the square points). In both Fig. 5(b) and 5(c), it is apparent that the DIH measured cumulative distributions agree with the expected number and volume distributions with both unbiased sampling and with settling bias. As the ML algorithm was not trained specifically on NIST beads, this demonstrates the robustness of the algorithm developed.

The dispersion of the NIST bead in the z-direction is approximately 8 cm, which is notably less than that observed in the spray cases. It is important to highlight that our model is trained using reconstruction planes. If a reconstruction plane captures a focused droplet, it is readily detected by the model. This model capitalizes on the features present in the reconstruction plane (in conjunction with the minimum intensity projection and the enhanced image) to identify the focused droplets. Unlike shadowgraphy or other conventional imaging methods, DIH can distinguish objects even if they are close to each other in z because the diffraction fringes of objects through the entire later plane, owing to the unique characteristics of compressive holographic imaging. Our model can perform well even under the conditions when a large number of particles are clustered within a narrow z-location, owing to the unique features of hologram formation. Thus, the z-span does not pose a challenge for our model.

*3.2 Spray Droplet Extraction & False Positive Rate Characterization*

**Table 2.** *Nozzle ID, reconstruction depth, extraction rate and false positive for different nozzles tested using the proposed model.*

| Nozzle ID | Minimum z (cm) | Maximum z (cm) | Depth (cm) | Extraction rate (%) * | | False positive (%)* |
|---|---|---|---|---|---|---|
| XR11003 | 14.0 | 39.0 | 25.0 | Center | 96.3 | 2.1 |
| | | | | Half-span | 94.8 | 2.1 |
| | | | | Edge | 93.7 | 2.8 |
| VF/F | 10.0 | 50.0 | 40.0 | 94.8 | | 2.8 |
| F/M | 4.0 | 45.0 | 41.0 | 94.5 | | 3.6 |
| M/C | 4.5 | 60.0 | 55.5 | 95.1 | | 4.8 |
| C/VC | 7.5 | 43.0 | 35.5 | 95.3 | | 3.8 |
| VC/XC | 10.0 | 50.5 | 40.5 | 93.5 | | 4.4 |
| XC/UC | 11.0 | 39.0 | 28.0 | 94.4 | | 3.5 |

As mentioned in the previous section, the model was trained using holograms of the XR11003 spray from the center, half-span, and edge locations. For testing, the ground truth was created by manually labeling each hologram using the same method mentioned for generating the training targets. When there is an overlap between the centroids of each signal in the prediction and the ground truth, they are treated as true positives. If there is no overlap, they are treated as ghost particles. The extraction rate is the percentage of true positives among the total number of droplets, while the false positive rate is defined as the percentage of ghost particles among the total number of droplets. For the test cases, we used a total of 5,000 droplets from each measurement location, corresponding to 23, 63 and 208 holograms for the center, half-span and edge. The proposed model performance with this spray is presented in Table 2, which shows extraction rates greater than 90% and false positive rates less than 5% across the three locations for this nozzle, demonstrating the accuracy of the proposed model after training.

With no additional training, the same test was repeated for six ASABE reference nozzles using 2,000 droplets from each center location. Based on the application of KL-Divergence to the extraction rates of XR11003, it was determined that a minimum of approximately 2000 droplets is required for a statistically stable extraction rate (see supplementary material). Therefore, for the reference nozzles, we used 2,000 droplets instead of the previously used 5000 in the case of XR11003. Manual labeling, used for testing, is a time-consuming process. Therefore, for the reference nozzles, the model's performance is tested using only the challenging situation, i.e., at the center location, where the number densities are high, resulting in overlapping interference patterns. The model's performance, reiterated in Table 2, indicates a consistent extraction rate exceeding 90% and a false-positive rate under 5% across the six different nozzles. This consistency underscores the model's general applicability in spray analysis. We observe that the machine learning algorithm is adept at identifying droplets with considerable accuracy, without an elevated rate of false positives. This accuracy is achieved without the necessity for a thresholding step and without confining the analysis to droplets within specific size ranges.

In Table 2, spray thickness along the z-direction varies between different nozzles. Minimum and maximum z values confine droplet encounter likelihood, but uniform application of these bounds can lead to increased computational time. For example, the M/C nozzle, with the largest depth, has droplets from $z=4.5$ cm to $z=60$ cm, necessitating analysis of 555 reconstruction planes (separated by 1000 µm) for precision. Conversely, the XR11003 nozzle exhibits a more confined droplet distribution from $z=14.0$ cm to $z=39.0$ cm, requiring analysis of only 251 planes and thus reducing computational effort. Adjusting z-boundaries per nozzle type optimizes computational efficiency without compromising model accuracy.

The models, including both U-Net and VGG16, have been trained to identify a dynamic range of droplet sizes emitted by six different nozzles, specifically ranging from 20-500 µm. In the context of the camera resolution used in this study, this range corresponds to approximately 1 to 28 pixels. To adapt the model for the detection of smaller droplets (those under 20 µm in diameter), a camera with a higher magnification imaging lens or a high-resolution sensor would be required. The speed or resolution of this technique is contingent upon the mode of acquisition utilized. In our current experiments, we employed a high-speed camera with a frame rate of 25 kHz. This allowed us to capture droplets moving at speeds of up to approximately 232 m/s. It should be noted, however, that the droplets were observed to move at significantly lower speeds. We maintained a high frame rate to ensure that individual droplets could be captured in multiple frames, aiding in the process of droplet tracking. Regarding the z-direction, the distance can be flexibly extended to any value situated between the camera and the collimator lens (as illustrated in Fig. 1). The models have been meticulously trained to segment features corresponding to a droplet present in the reconstruction plane, upon detection.

*3.3 Spray Visualization*

Fig. 6 shows a qualitative representation of the ML processed results, specifically this figure contains a shadowgraph of the XC/UC spray along with a 3D rendering of droplet fields from the three DIH measurement locations, rendered using OVITO (Stukowski 2009). We have added a supplementary video that includes the shadowgraphy and the rendered DIH videos for the three locations. Holograms were collected at 60 cm downstream from the nozzle exit, i.e., hologram image locations are farther downstream than depicted in the Fig. 6. From the shadowgraph is evident that complete atomization occurs a few centimeters downstream from the nozzle and the droplets reaching the DIH measurement plane are largely spherical in shape and their distribution

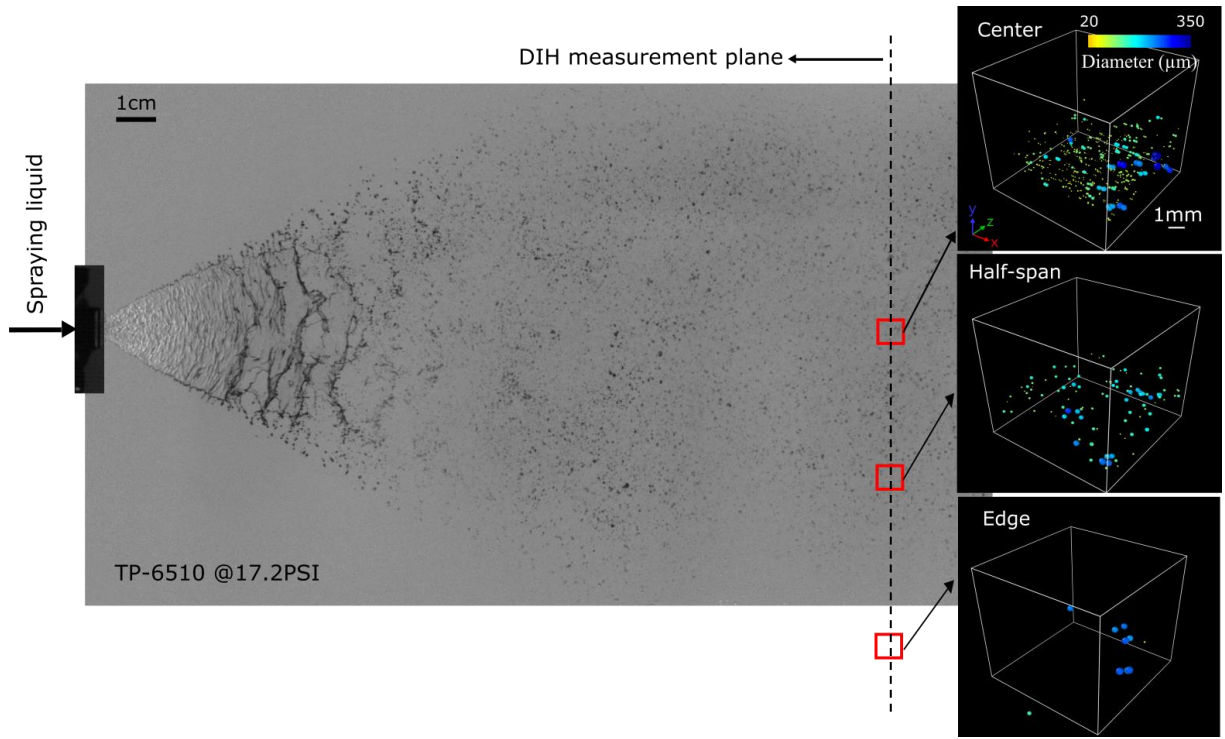

*Fig. 6. A shadowgraph image and the corresponding 3D droplet fields for the center, half-span, and edge locations at the DIH measurement plane of the reference nozzle VC/XC (TP6510). The rendered droplet sizes are doubled (for all droplets) in the DIH visualization for improved clarity.*

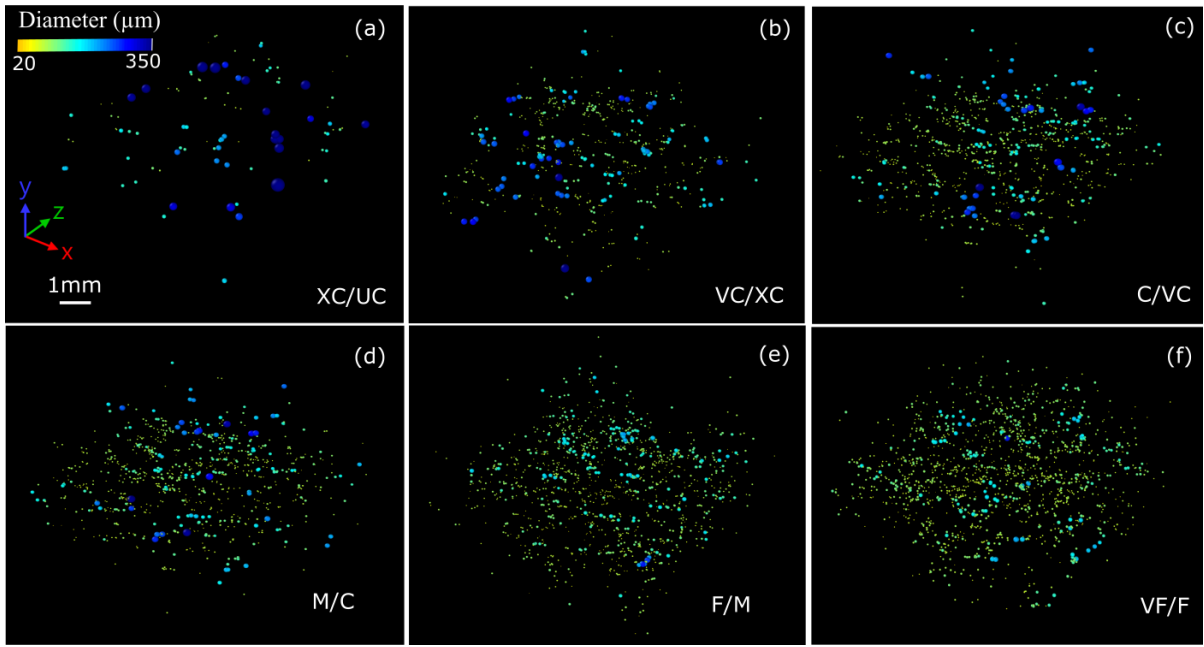

*Fig. 7. Reconstructed snapshots of 3D droplet distribution from center regions of sprays 60 cm downwind of the nozzle exit for ASABE reference nozzles. (a) TP6515-SS, (b) TP6510-SS, (c) TP8008-SS, (d) TP11006-SS, (e) TP11003-SS, and (f) TP11001-SS*

is no longer evolving. In the DIH renderings the number density of droplets reduces from the center to the edge region, as is typical for flat fan sprays. The droplets in edge regions are relatively larger than that found in the center and half-span locations, which is also expected (Ellis et al. 1997). Fig. 7(a)-7(e) show renderings of the center region of the six different reference nozzle sprays, corresponding videos showing over two milliseconds of measurement are provided in the supporting information. A colormap is used to represent the droplet sizes, where yellow and blue represent the smallest and largest droplets, respectively. Consistent with the ASABE standards, the larger droplets (in blue) are easily visible in Fig. 7(a-d) but are less present in Fig. 7(e) and Fig. 7(f), i.e., these droplets are produced by the F/M and VF/F nozzles at a much lower rate. Correspondingly, the fraction of small droplets (in yellow) also substantially increases for these nozzles.

*3.4 DIH Derived Spray Properties*

Droplet number and volumetric distributions were constructed from a 1250 hologram subsample of 25000 consecutive frames and are shown in Fig. 8 for the center region of all six reference nozzles. Due to the resolution of the camera the minimum resolvable diameter from these holograms is 18.2 µm. Fig. 8(a) and 8(b) show the cumulative number and cumulative volume distributions respectively as determined by binning the raw data with a bin size of 5 µm. A kernel density estimator with a Gaussian kernel bandwidth of 0.05 was used to calculate the number-based droplet size distribution (DSD) functions of the of the six reference nozzles (shown in Fig. 8c). These droplet size distributions resemble unimodal, nearly lognormal with the peak value occurring in the 30-40 µm range for all nozzles. A magnified view of the tail region, presented as an inset, unveils that despite having similar mode values, the coarser nozzles indeed generate larger droplets identifiable in the DIH. This is further exemplified in the volumetric droplet distributions (VDSDs), which are presented in Fig. 8(d). These VDSDs are smoothed using a moving average with a window size of 5 µm, resulting in the cut off of the VDSDs at ~400 µm. Similar to the number-based distributions, the VDSDs are nearly lognormal.

From the DSDs, the Sauter mean diameter $D_{32}$ is estimated as $D_{32} = \frac{\sum_{i=1}^{n} N_i d_i^3}{\sum_{i=1}^{n} N_i d_i^2}$, where $N_i$ is the number of droplets corresponding to the diameter bin centered around $d_i$. We remark that because DIH enables detection of individually droplets, $D_{32}$ is calculated using this discrete form, as opposed to from integral of the third and second moments of the number-based distributions. $D_{32}$ variation along the six different nozzle categories and for the three different measurement locations is shown in Fig. 9(a). As expected, $D_{32}$ decreases systematically as the nozzle class decreases from XC/UC to VF/F. For all nozzles, $D_{32}$ also increases while moving from the center to the edge region. This is consistent with prior spray measurements for traditional techniques, for pure water sprays (Li et al. 2021). and has been rationalized through examination of disparate sheet breakup modes on the edge of a spray in comparison to the center region (Kooij et al. 2018).

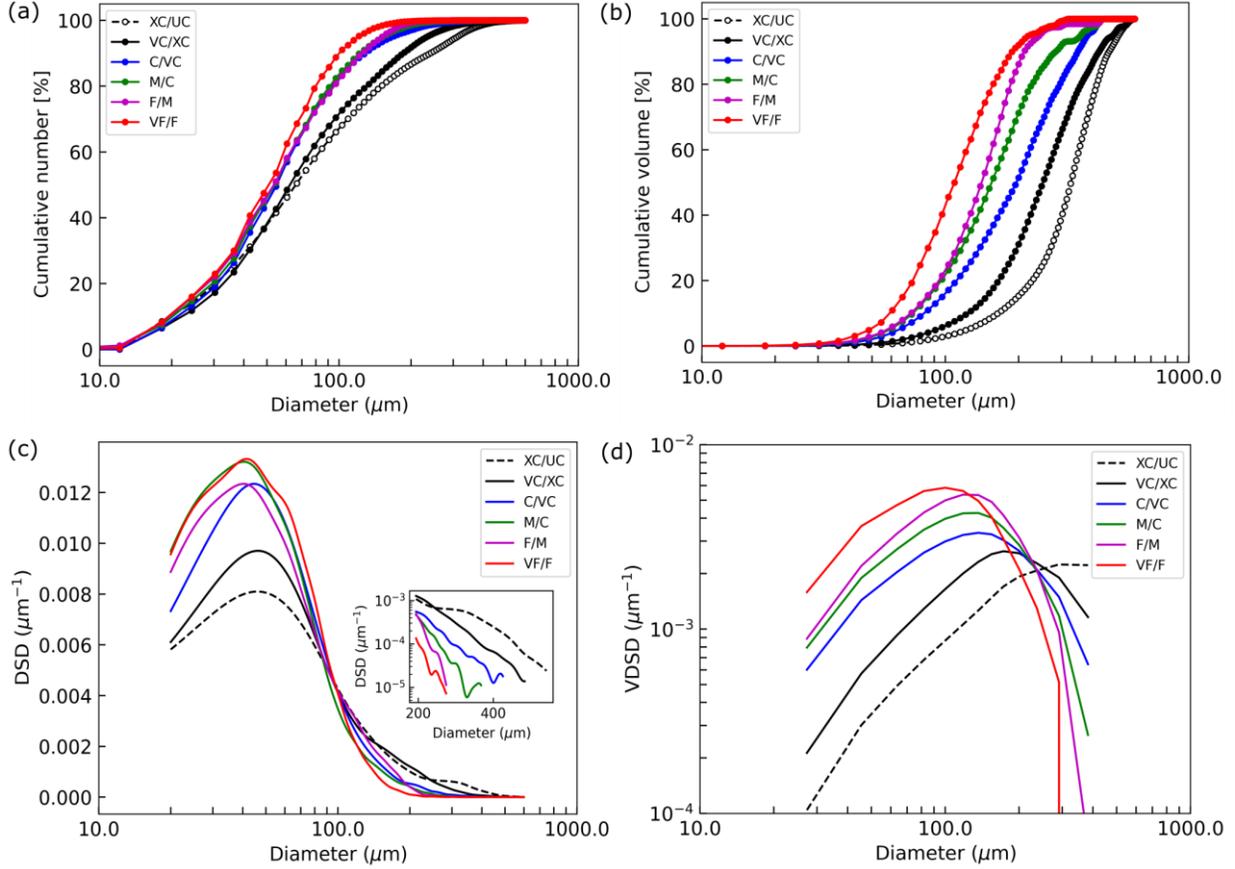

*Fig. 8.* *Cumulative number (a) and volume (b) distributions for the six reference nozzles, extracted from DIH measurements in the center region. The corresponding cumulative number (c) and volume (d) distributions are shown in the lower panel.*

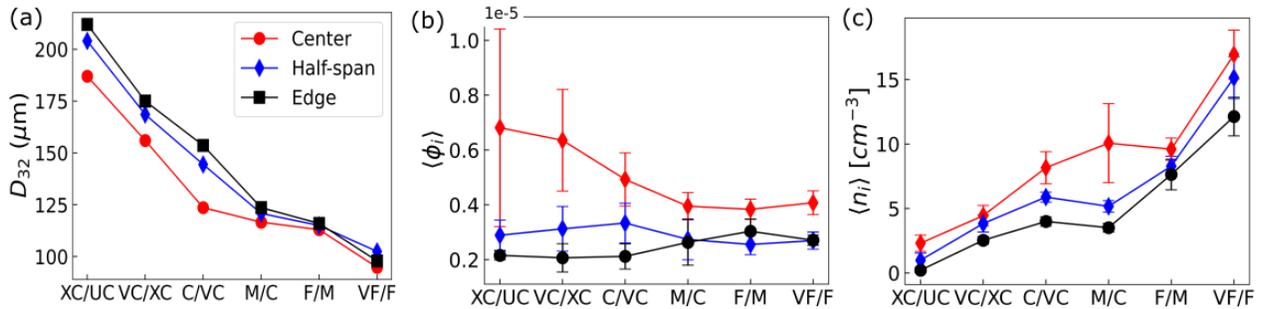

*Fig. 9.* *(a) Variation of Sauter mean diameter ($D_{32}$) at the center, half-span and edge regions for the six different reference nozzles. Average volume fraction $\langle \phi_i \rangle$ (b) and average number concentration n (c) across the six different reference nozzles. The vertical error bars represent the standard deviation across 500 subsampled holograms.*

Because the viewing volume is well-defined, DIH also enables inference of the volume fraction of droplets at each measurement location. The volume fraction in the $i^{th}$ frame is estimated as $\phi_i = \left( \frac{V_i}{x_{rec} \times y_{rec} \times z_{rec}} \right)$, where $V_i$ represents the total droplet volume extracted in the $i^{th}$ frame. It is important to note that at the measurement location, all particles maintain a spherical shape as they are far downstream from the nozzle exit. This consistency in shape allows for an accurate measurement of the droplets' volume using the predicted diameter. The values $x_{r_{ec}}$, $y_{r_{ec}}$, and $z_{r_{ec}}$ correspond to the sensor width, height, and reconstruction depth of the hologram, respectively. While $x_{r_{ec}}$, $y_{r_{ec}}$ remain the same for all reference nozzles, $z_{r_{ec}}$ varies depending on the reconstruction depths used for each reference nozzle, as indicated in Table 2. The estimated value of $z_{r_{ec}}$ is calculated as the difference between $z_{max}$ and $z_{min}$, where $z_{max}$ and $z_{min}$ refer to the maximum and minimum longitudinal locations, respectively, that contain droplets in the prediction. The variation in the average volume fraction $\langle \phi_i \rangle$, is estimated using 500 subsampled holograms for each nozzle, and results are plotted in Fig. 9(b). The vertical error bars represent the standard deviation in volume fraction $\phi_i$ across the 500 subsample holograms. For all cases, the average volume fraction is less than $O(10^{-5})$, suggesting that the sprays are in the one-way coupling regime, where droplets do not significantly influence the background flow characteristics, and the likelihood of droplet collisions is low. In the center region where the spray is finer, the average volume fraction decreases due to a reduction in the proportion of larger droplets.

Similar to the volume fraction, the average number concentration $\langle n_i \rangle$ is also estimated for all the reference nozzles and plotted in Fig. 9(c). For the majority of the reference nozzles, even though $D_{32}$ increases towards the edge location, $\langle \phi_i \rangle$ reduces (Fig. 9b). This reduction is attributed to the decrease in $\langle n_i \rangle$ (shown in Fig. 9c) and an observed increase in $z_{r_{ec}}$ as we move away from the center location. While the variability in $\langle \phi_i \rangle$ with holograms, as well as locations, decreases when moving from coarse to fine spray categories (as shown in Fig. 9b), the variability in $\langle n_i \rangle$ (Fig. 9c) increases. This suggests that larger variations in the number of smaller droplets are not well captured by volume fraction estimates. Therefore, estimates such as $\langle n_i \rangle$ can be a better measure for drift analysis compared to volume fraction.

Beyond size distributions, because of the ability to identify individual droplets, DIH enables determination of quantities not typically reported for sprays. For example, mass and momentum fluxes of droplets can be readily determined. Fig. 10 presents the results for the six different nozzles. For these calculations, instead of using subsampled images, consecutive images for a duration of one second were employed for flux estimations. The mass flux was determined by calculating the total droplet mass passing through the YZ plane (see Fig. 10a). To calculate the momentum flux, individual droplet velocities are estimated by tracking droplets across consecutive frames. In each consecutive hologram, droplets are uniquely identified using the nearest neighbour algorithm (Kumar et al. 2019a). If there are missing droplets in consecutive frames or if multiple droplets are detected as separate entities but belong to the same droplet, the trajectories are merged using velocity kinematics. This process ensures that the missing droplets are accounted for, and any incorrectly detected separate droplets are properly merged to maintain the continuity of the droplet trajectories. Fig. 10b illustrates the mass and momentum flux estimations for the six different spray categories. The reduction in mass and momentum flux as the spray becomes finer can be attributed to the increased presence of smaller droplets.

Validation of the mass flux estimation is achieved using NIST bead measurements as well. In these measurements, the mass flux is manually determined. With 2.5g of NIST beads poured through the funnel in a total time of 30 seconds, the mass flux for the funnel's cross-sectional area is approximated at 1.31 kg/m$^2$s. A mass flux of 1.35±0.05 kg/m$^2$s is estimated from DIH measurements, with the uncertainty derived from five different time instants, each comprising 200 consecutive holograms. Agreement within the uncertainties of the measurements is observed, demonstrating the reliability of the mass flux estimation. Discrepancies between the original and estimated mass flux are attributed to factors including cross-sectional area mismatch and variations in the bead dispersion during measurement.

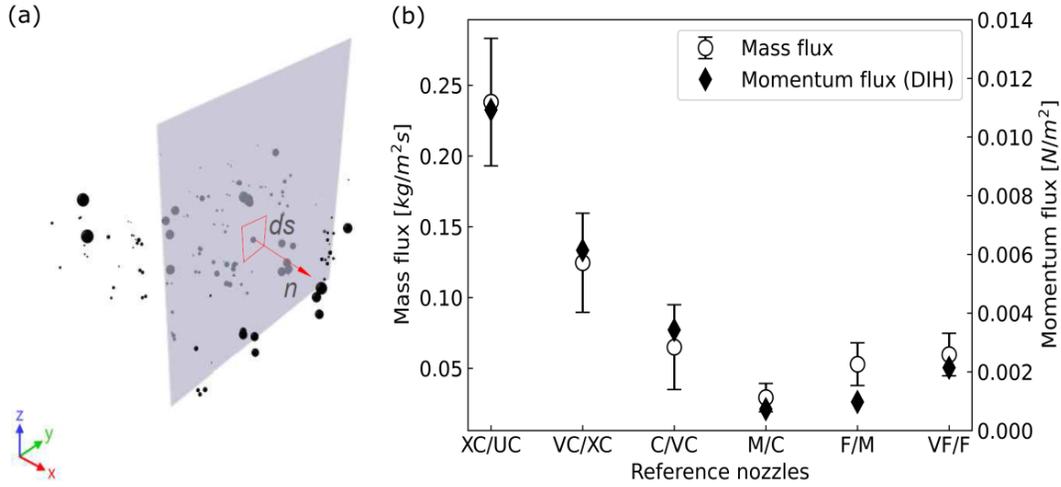

*Fig. 10.* *(a) Schematic diagram illustrating the plane used for the flux estimation and (b) mass and momentum flux across the six different tested nozzles in the center region. Measurement using consecutive holograms from 5 non-overlapping time instant is used for the error bar estimation.*

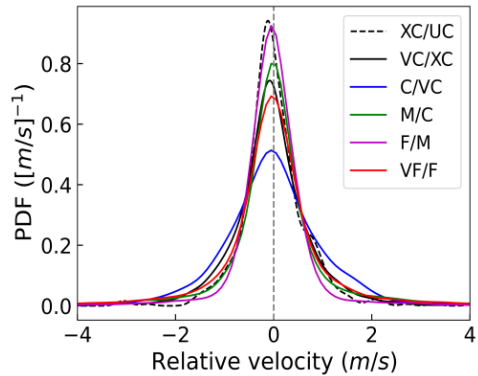

*Fig. 11.* *Distribution of radial relative velocities of droplet pairs across all droplet size classes*

Individual droplet velocity tracking also enables the determination of relative droplet velocities, which is of considerable utility in calculating droplet collision rate coefficients (Kumar et al. 2019a). From the DIH renderings, the radial relative velocities between droplets are estimated as $(v_1 - v_2) \cdot \mathbf{r}/|\mathbf{r}|$. $\mathbf{r}$ is the separation vector that goes from droplet 2 to 1 and the absolute coordinate of the relative velocity aligns with the separation vector. $v_1$ and $v_2$ are the droplet velocities obtained through tracking (Kumar et al. 2023). The distance covered between each frame is estimated from the droplet tracking as mentioned previously. The corresponding velocity is gauged using the camera frame rate used for the current study (25 kHz). The average velocity across all the detected frames is used as the droplet velocity $v$ for the relative velocity estimations. Negative and positive relative velocities indicate droplets moving towards and away from each other, respectively. For this analysis, the maximum separation distance considered was 1 mm, as a larger value was found to have no significant influence on the final measurement. Fig. 11 presents the probability density functions of relative velocity for all droplet size classes. The relative velocity distribution peaks are consistently near zero relative velocity. In the C/VC and VC/XC spray categories, there is a relatively large fraction of droplet pairs with larger relative velocities in the range [-1, -2 m/s] (Fig. 11). Notably, the distribution for VC/XC shows a slightly positive skew, indicating that the majority of droplet pairs in this category are converging. In the XC/UC category, most pairs exhibit small relative velocities, which can be attributed to the low number density and large axial momentum that reduce the longitudinal movement of the droplets. To our knowledge, relative droplet velocities in agricultural sprays have not been examined previously, and we suggest that use of DIH enables greater insight into droplet dynamics following breakup.

*3.5 Comparison to Laser Diffraction*

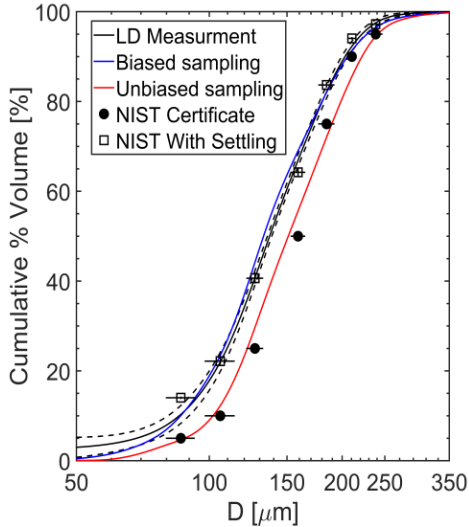

*Fig. 12. Comparison with Laser Diffraction (LD) inferred cumulative volume distributions estimated from the LD and DIH based NIST measurements.*

Previous studies (Privitera et al., 2023) have highlighted that measurements obtained using different techniques can be influenced by the droplet size distributions. Additionally, Sijs et al. (2021) demonstrated that variations in results are more pronounced when larger droplets are involved. This suggests that measurements conducted across different nozzles with varying droplet size distributions or at different locations within the same spray field can lead to biased results based on the local droplet size distribution. The utilization of DIH presents a potential solution to overcome the limitations associated with conventional techniques when dealing with diverse, spatially variable droplet populations. Laser diffraction (LD) systems are commonly used to characterize agricultural sprays (ISO 2018, ASABE 2020, Fritz et al., 2011). Fig. 12 illustrates the performance of the LD system in conjunction with DIH using the proposed model, based on a NIST traceable microbead measurement (the results of Fig. 5). LD systems are often specifically employed to capture the cumulative volume distribution, which is a commonly utilized metric in spray analysis. However, as described by Fredericks (2020) and Kumar et al., (2020) LD measurements are biased towards droplets moving at slower speeds, increasing the dwell time within the measurement volume. The bias becomes evident when examining the calculated cumulative volume distribution obtained through Laser Diffraction (LD) for settling beads. LD systems tend to retrieve only a skewed cumulative volume distribution. On the other hand, DIH yields a biased distribution only when biased sampling methods like consecutive frames are used. Interestingly, by avoiding consecutive frames, the original distribution, which has been validated by the National Institute of Standards and Technology (NIST), is restored. This approach is detailed in the subsection titled "Model Validation: Bead Measurements". In terms of D0.5, the NIST with settling has a D0.5 around 139 µm, while the LD measurements and DIH (ML)-based biased sampling measured 136 µm and 133 µm, respectively. The original NIST D0.5 is 155 µm, and the DIH (ML)-based unbiased sampling measured a D0.5 of approximately 153 µm. The close match between the D0.5 of unbiased sampling and the NIST original data clearly shows the effectiveness of the proposed ML-based DIH compared to the LD system.

## 4. Discussions
*4.1 Advantage of ML-based DIH processing over the conventional DIH processing*

The segmentation methods used in non-ML DIH often require human intervention for fine-tuning parameters to achieve optimal performance, particularly when dealing with holograms captured under varying conditions (Kumar et al. 2019b, Kumar et al. 2020, Li et al. 2021, Gao et al. 2013, Wang et al. 2022). This necessity for manual adjustment limits the generalizability of the method used. The inherent unsteadiness in spray systems results in variations in number density between frames. In such contexts, every individual hologram mandate distinct parameter adjustment when using the conventional method. On the other hand, semantic segmentation models, like U-Net are learning complex, hierarchical features and relationship in the data to segment and classify each pixel effectively. U-Net learns a hierarchy of features through its layers of neurons. Each layer learns to recognize more complex patterns by combining the similar features learned in the previous layers. The extracted features are then classifying each pixel into different categories (for example droplet and background in this study). U-Net uses ReLU, which introduces non-linear pattern detection, which is not possible in conventional methods.

Furthermore, to mitigate noise interference in conventional DIH methods, particles with a pixel number exceeding a certain threshold value are typically considered as particles in previous studies (Wange et al. 2022). This can inadvertently filter out weak signals, including those from very small particles. The proposed model avoids filtering out weak signals and can detect particles as small as 1 pixel, which corresponds to ~20 µm. Using a magnification lens or a camera with higher resolution can help detect smaller droplets using the same model.

Our model has been specifically developed for spray visualization in agricultural applications. It is noteworthy to mention that our training solely relied on holograms from the XR11003 nozzle. Despite this, the model exhibits commendable performance on spray nozzles distinct from XR11003, achieving an extraction rate of over 90% and a false positive rate of less than 5%. The feature extraction-based detection of the ML method allows the use of the same model for different optical settings as well as different spray measurements. On the other hand, using the conventional DIH method involves parameter tuning for each frame, making it difficult and time-consuming to process a large number of holograms. For example, approximately 200 holograms need to be processed to obtain a statistically stable result for the XR11003 nozzle. Meanwhile, ML-based DIH processing facilitates automated batch processing for a very large number of holograms acquired under different conditions. To further enhance the model's adaptability across different dynamic ranges, we believe that generating a synthetic training set or augmenting the current set with additional sample holograms could be beneficial.

*4.2 Application scenario of the technology*

Characterization of agricultural sprays is crucial in the development of agricultural nozzles and in assessing the performance of formulated products used in tank mixes (Makhnenko et al. 2021). While LD and other 1D- temporal measurement techniques are currently widely used for spray characterization, DIH can be used as an alternate method. The incorporation of the proposed ML method enhances both the speed and generalizability of DIH processing, enabling the spray characterization without human intervention. Additionally, it facilitates the measurement of parameters that are challenging to ascertain with conventional spray measurement techniques, such as spatial and temporal variation in DSDs, mass flux, and relative velocity. This comprehensive characterization of agricultural sprays can be used to better predict field performance of products during development. The applicability of DIH and ML-based processing extends beyond the specific settings of lab experiments. The compact setup of DIH could enable development for its application in diverse settings, including in-situ measurements in canopy near plant surfaces, potentially characterizing droplet-leaf interactions such as splash or bounce, enabling the customization of application specific droplet characteristics (Qiu et al., 2022).

## 5. Conclusion

We introduce a novel machine learning (ML) based Digital In-line Holography (DIH) methodology for the analysis of pesticide sprays in agricultural applications. Our approach combines the strengths of two neural networks, creating a robust and efficient system for spray data interpretation. The primary network is designed to segment the 3D droplet field from the numerically reconstructed optical field, while the secondary network acts as a binary classifier, effectively reducing segmentation noise. Our methodology has undergone rigorous validation using standard NIST traceable glass beads, with sizes ranging from 50 to 350 µm, and across six distinct nozzle types that generate a wide variety of sprays. This ensures our approach is applicable to the full spectrum of droplet sizes typically found in agricultural sprays. The application of our ML-based DIH methodology has consistently achieved an extraction rate surpassing 90% and false positives kept under 5% across the whole range of droplet number densities and diameters tested in the study. Furthermore, we have demonstrated the ability of our approach to obtain both 3D size distribution and velocity of droplets, which enables the estimation of mass and momentum flux at different locations and the calculation of relative velocities of droplet pairs. These measurements, which are typically challenging to capture using conventional spray measurement techniques such as Laser Diffraction (LD) and Phase Doppler Particle Analyzer (PDPA), are vital for comprehending spray behavior and refining spray applications in agricultural operations.

While our approach has demonstrated significant promise, like all innovative methodologies, it has limitations that provide avenues for future exploration and enhancement. A key limitation is the processing speed. Despite its high accuracy, our approach is not yet capable of real-time processing, with the U-Net used for hologram segmentation currently consuming over 80% of the processing time. Future work could focus on refining the ML architecture and harnessing advancements in GPU processing speed to overcome this challenge and potentially achieve real-time processing capabilities. Additionally, the miniaturization of the DIH system for field deployment, particularly for integration with UAVs, is another area for improvement (Bristow et al., 2023). This would make our approach more practical and accessible for a wider range of agricultural operations, enabling measurements in field settings, such as spray characterization and droplet transport at the point of application.

**Reference:**
1. ASABE 2020 Spray Nozzle Classification by Droplet Spectra. ANSI/ASAE S572.3